\documentclass[preprint,showpacs,preprintnumbers,showkeys,amsmath,amssymb]{revtex4}
\usepackage{epsfig}
\usepackage{dcolumn}
\usepackage{bm}

\begin{document}

\title{A new comprehensive study of the 3D random-field Ising model via
sampling the density of states in dominant energy subspaces}

\author{Nikolaos G. Fytas}
\altaffiliation[]{Corresponding author: nfytas@phys.uoa.gr}
\author{Anastasios Malakis}
\affiliation{Department of Physics, Section of Solid State
Physics, University of Athens, Panepistimiopolis, GR 15784
Zografos, Athens, Greece}


\begin{abstract}
The three-dimensional bimodal random-field Ising model is studied
via a new finite temperature numerical approach. The methods of
Wang-Landau sampling and broad histogram are implemented in a
unified algorithm by using the N-fold version of the Wang-Landau
algorithm. The simulations are performed in dominant energy
subspaces, determined by the recently developed critical minimum
energy subspace technique. The random-fields are obtained from a
bimodal distribution, that is we consider the discrete
$(\pm\Delta)$ case and the model is studied on cubic lattices with
sizes $4\leq L \leq 20$. In order to extract information for the
relevant probability distributions of the specific heat and
susceptibility peaks, large samples of random-field realizations
are generated. The general aspects of the model's scaling behavior
are discussed and the process of averaging finite-size anomalies
in random systems is re-examined under the prism of the lack of
self-averaging of the specific heat and susceptibility of the
model.
\end{abstract}

 \maketitle

\section{Introduction}

The random-field Ising model (RFIM)~\cite{imry75} is one of the
most studied glassy magnetic
models~\cite{imbrie84,bricmont87,belanger98}. The 3D RFIM consists
of Ising spins $S_{i}$ on a simple cubic lattice, governed by the
Hamiltonian:
\begin{equation}
\label{eq:1}
\mathcal{H}=-J\sum_{<i,j>}S_{i}S_{j}-\sum_{i}h_{i}S_{i}
\end{equation}
where $J>0$ is the interaction constant and $h_{i}$ are quenched
random-fields, obtained from a bimodal distribution
$P(h_{i})=\frac{1}{2}
\delta(h_{i}-\Delta)+\frac{1}{2}\delta(h_{i}+\Delta)$. $\Delta$
denotes the disorder strength, also called randomness of the
model. Although nowadays it is believed that the phase transition
from the ordered to the disordered phase of the model is of
second-order, a complete set of critical exponents fulfilling a
widely accepted set of scaling relations has not been established.
In fact, there is a strong disagreement in literature concerning
the overall thermal and magnetic behavior of the
model~\cite{middleton02,hartmann01}. This may be due to a mistaken
comprehension of some theoretical concepts in random systems, such
as the concept of averaging that will be discussed below.

The rest of the paper is laid out as follows.  In
Section~\ref{section2} we present the numerical schemes utilized
for the study of the RFIM. The process of averaging finite-size
anomalies in random systems and the significance and implications
of the non trivial property of the lack of self-averaging of the
specific heat and susceptibility of the model, are discussed in
Section~\ref{section3}. Finally, we summarize in
Section~\ref{section4}.

\section{Numerical techniques}
\label{section2}

Numerically the RFIM has been approached using
traditional~\cite{rieger93,rieger95} but also more sophisticated
Monte Carlo techniques~\cite{newman96}. However, the nature of the
model demands enormous computer resources. Furthermore, in order
to get a good estimate of the mean properties of the system, it is
necessary to repeat the simulations for a large number of
realizations of the random-fields. Here, the numerical procedure
concentrates on the determination of the density of states (DOS)
$G(E)$ of the model and on the corresponding thermodynamic
quantities.

For the application of the Wang-Landau (WL)
algorithm~\cite{wang01} in a multi-range approach we follow the
N-fold description of Schulz \textit{et al.}~\cite{schulz03}. The
random walk is not allowed to move outside of any particular
subrange, and we always increment the histogram $H(E)\rightarrow
H(E)+1$ and the DOS $G(E)\rightarrow G(E)*f_{j}$ after a spin-flip
trial. Here, $f_{j}$ is the value of the WL modification factor
$f$~\cite{wang01} at the jth iteration, in the process
$(f\rightarrow f^{1/2})$ of reducing its value to 1, where the
detailed balance condition is satisfied. In all our simulations
the control parameter takes the initial value:
$f_{j=1}=e\approx2.71828...$, while when starting a new iteration
it is changed according to the sequence $f_{j+1}=\sqrt{f_{j}},
j=1,2,...,20$~\cite{wang01,malakis04}. For the histogram flatness
criterion we use a flatness level of $0.05$. The accumulation of
numerical data for the application of the broad histogram (BH)
method of Oliveira \textit{et al.}~\cite{oliveira96} and also the
updating of appropriate $(E,M)$ histograms is carried out in the
final stage of the WL process, by using the N-fold iteration
$j=12-20$~\cite{malakis05}. The approximation of the DOS, in the
last WL iteration, $G_{WL}(E)$, and the high-level $(j\gg1)$ WL
$(E,M)$ histograms, $H_{WL}(E,M)$, are then used to estimate the
magnetic properties in a temperature range, which is covered, by
the restricted energy subspace $(E_{1},E_{2})$ as:
\begin{equation}
\label{eq:2} \langle M^{n}\rangle=\frac{\sum_{E}\langle
M^{n}\rangle_{E}G(E)e^{-\beta E}}{\sum_{E}G(E)e^{-\beta E}}\cong
\frac{\sum_{E\in(E_{1},E_{2})}\langle
M^{n}\rangle_{E,WL}G_{WL}(E)e^{-\beta
E}}{\sum_{E\in(E_{1},E_{2})}G_{WL}(E)e^{-\beta E}}
\end{equation}
The microcanonical averages $\langle M^{n}\rangle_{E}$ are
obtained from the $H_{WL}(E,M)$ histograms as:
\begin{eqnarray}
\label{eq:3}\langle M^{n}\rangle_{E}\cong\langle
M^{n}\rangle_{E,WL}\equiv
\sum_{M}M^{n}\frac{H_{WL}(E,M)}{H_{WL}(E)}\nonumber
\\ H_{WL}(E)=\sum_{M}H_{WL}(E,M)
\end{eqnarray}
and the summation in $M$ runs over all values generated in the
restricted energy subspace $(E_{1},E_{2})$~\cite{malakis05}.
Similarly we obtain the microcanonical estimators necessary for
the application of the BH method, using the well-known broad
histogram equation~\cite{oliveira96}:
\begin{equation}
\label{eq:4} G(E)\langle N(E,E+\Delta E_{n})\rangle_{E}=G(E+\Delta
E_{n})\langle N(E+\Delta E_{n},E)\rangle_{E+\Delta E_{n}}
\end{equation}
where $N(E,E+\Delta E_{n})$ is the number of possible spin flip
moves from a microstate of energy $E$ to a microstate with energy
$E+\Delta E_{n}$, which are known during the N-fold process.

For a particular random-field realization the specific heat and
its peak are easily obtained with the help of the usual
statistical sums. The critical minimum energy subspace (CrMES)
scheme~\cite{malakis04,malakis05} uses only a small but dominant
part $(\widetilde{E}_{-},\widetilde{E}_{+})$ of the total energy
space $(E_{min},E_{max})$ to determine the specific heat peaks.
Let $\widetilde{E}$ denotes the value of energy producing the
maximum term in the partition function at the pseudocritical
temperature (corresponding to the specific heat peak) and
$S(E)=\ln{G(E)}$ the microcanonical entropy. Then the CrMES
approximation is defined by the following equations:
\begin{equation}
\label{eq:5}
C_{L}(\widetilde{E}_{-},\widetilde{E}_{+})=N^{-1}T^{-2}\left\{\widetilde{Z}^{-1}
\sum_{\widetilde{E}_{-}}^{\widetilde{E}_{+}}E^{2}\exp{[\widetilde{\Phi}(E)]}-
\left(\widetilde{Z}^{-1}\sum_{\widetilde{E}_{-}}^{\widetilde{E}_{+}}E
\exp{[\widetilde{\Phi}(E)]}\right)^{2}\right\}
\end{equation}
\begin{equation}
\label{eq:6} \widetilde{\Phi}(E)=[S(E)-\beta
E]-\left[S(\widetilde{E})-\beta
\widetilde{E}\right],\;\;\widetilde{Z}=\sum_{\widetilde{E}_{-}}^{\widetilde{E}_{+}}\exp{[\widetilde{\Phi}(E)]}
\end{equation}
where $(\widetilde{E}_{-},\widetilde{E}_{+})$ is the minimum
dominant subrange, satisfying the following accuracy criterion:
\begin{equation}
\label{eq:7}
\left|\frac{C_{L}(\widetilde{E}_{-},\widetilde{E}_{+})}{C_{L}(E_{min},E_{max})}-1\right|\leq\textit{r}
\end{equation}
with $r=10^{-6}$. Note that, the above accuracy is extremely
demanding compared to the statistical errors produced by the DOS
method (i.e. the WL method) and to the large sample-to-sample
fluctuations of the RFIM that will be discussed below in
Section~\ref{section3}.

Using an ensemble of macroscopic samples of size $L$ corresponding
to different random-field realizations we have applied the
described scheme in a broad energy(magnetization) space (total
CrME(M)S of the ensemble) that covers the overlap of the dominant
energy(magnetization) subspaces for all realizations of the
ensemble. This practice has the advantage that the approximation
of the specific heat and susceptibility for a particular
random-field is accurate in a wide temperature range, including
its pseudocritical temperature. Despite the strong fluctuations of
the energy value corresponding to the maximum term of the
partition function $Z$, the union of the CrME(M)S for large
samples of random-fields is, in any case, a quite small subspace.

\section{Averaging finite-size anomalies. Lack of self-averaging}
\label{section3}

For a disordered system one has to perform two distinct kinds of
averaging. Firstly, for each random-field realization the usual
thermal average has to be carried out and secondly one must
average over the distribution of the random parameters. The latter
makes it clear that large ensembles of random-fields must be
generated in order to estimate properly the mean properties of the
system. Following the methods described above in
Section~\ref{section2} the thermal average for the specific heat
is given by Eq.~(\ref{eq:5}), while the susceptibility $\chi$
reads as:
\begin{equation}
\label{eq:8} \chi=\frac{N}{T}\left\{\langle M^{2}\rangle-\langle
M\rangle^{2}\right\}
\end{equation}
with $N=L^{3}$. Let $C_{m}(T)$ and $\chi_{m}(T)$ denote the
specific heat and susceptibility of a particular random-field
realization $m$ in an ensemble of $M$ realizations
$(m=1,2,...,M)$. The corresponding pseudocritical temperatures
$T_{L}^{\ast}(C_{m}(T))$ and $T_{L}^{\ast}(\chi_{m}(T))$ depend on
the particular realization of the random-field and for large
values of the randomness $\Delta$, they are strongly fluctuating
quantities. The locations of the specific heat and susceptibility
peaks may be then denoted by $(C_{m}^{\ast},T_{L,C;m}^{\ast})$ and
$(\chi_{m}^{\ast},T_{L,\chi;m}^{\ast})$, respectively.

In previous studies~\cite{rieger93,rieger95}, the averaging
process over a large number of random-fields has been carried out
on the averaged curve of the specific heat or susceptibility,
without first raising the question of whether this averaged curve
is the proper statistical representative of the system.
Specifically, the following sample averages have been considered
for the specific heat and susceptibility~\cite{rieger93,rieger95}:
\begin{equation}
\label{eq:9}
[C]_{av}=\frac{1}{M}\sum_{m=1}^{M}C_{m}(T);\;\;[\chi]_{av}=\frac{1}{M}\sum_{m=1}^{M}\chi_{m}(T)
\end{equation}
The finite-size scaling behavior of the peak of these averaged
curves was then studied by assuming that the maxima
$[C]^{\ast}_{av}=([C]_{av})^{\ast}$ and
$[\chi]^{\ast}_{av}=([\chi]_{av})^{\ast}$ obey power laws (for
details see Refs.~\cite{rieger93,rieger95}). Note that, these
averaged curves ($[C]_{av}$ and $[\chi]_{av}$) are very sensitive
to the property of self-averaging due to the fact that the
corresponding thermodynamic quantities are characterized by broad
distributions in the thermodynamic limit.
\begin{figure}[htbp]
\includegraphics*[width=10 cm]{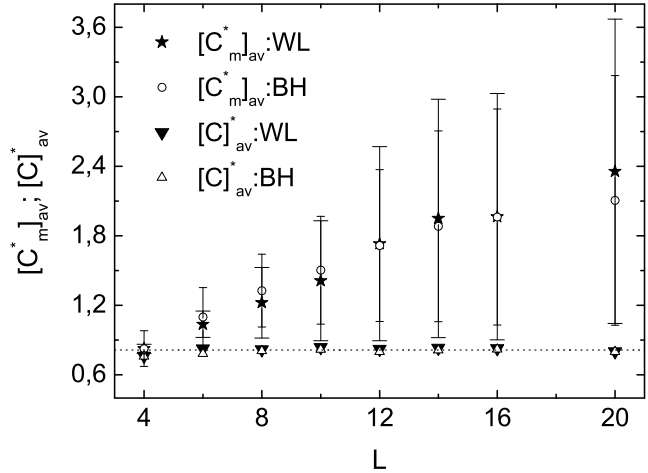}
\caption{\label{fig1}Finite-size behavior of the averages
$[C_{m}^{\ast}]_{av}$ and $[C]_{av}^{\ast}$ for the case
$\Delta=2$, for both the WL and BH methods used. The error bars
represent the sample-to-sample fluctuations (see text). The
behavior of $[C]_{av}^{\ast}$ appears as a random fluctuation
around the value $0.815$, as shown by the dotted line.}
\end{figure}

\begin{figure}[htbp]
\includegraphics*[width=10 cm]{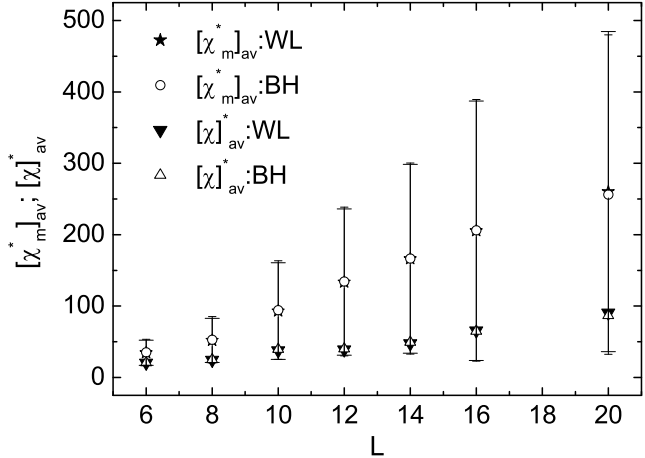}
\caption{\label{fig2}Finite-size behavior of the averages
$[\chi_{m}^{\ast}]_{av}$ and $[\chi]_{av}^{\ast}$ for the WL and
BH methods, also for the case $\Delta=2$.}
\end{figure}

In this work, in addition to the above averaging expressions, we
study the sample-averages of the individual specific heat and
susceptibility maxima, defined by:
\begin{equation}
\label{eq:10}
[C_{m}^{\ast}]_{av}=\frac{1}{M}\sum_{m=1}^{M}C_{m}^{\ast};\;\;[\chi_{m}^{\ast}]_{av}=\frac{1}{M}\sum_{m=1}^{M}\chi_{m}^{\ast}
\end{equation}
These mean values, together with the corresponding peaks of the
averaged curves of Eq.~(\ref{eq:9}), are shown in
Figs.~\ref{fig1},\ref{fig2}. In our simulations we used an
ensemble of $M=1000$ random-field realizations for $L\leq 12$ and
$M=500$ for $L=14-20$. To quantify the sample-to-sample
fluctuations of the specific heat (susceptibility) peaks we define
the standard deviation of $C_{m}^{\ast}$ ($\chi_{m}^{\ast}$), over
a sample of $M$ random-field realizations as
$\sigma(C_{m}^{\ast})$ ($\sigma(\chi_{m}^{\ast})$). This important
parameter will be illustrated in our figures as error bars, but
should not be in any case confused with the existing statistical
errors.

Fig.~\ref{fig1} illustrates the finite-size behavior of the peaks
of the average $[C_{m}^{\ast}]_{av}$ and that of the averaged
curve $[C]_{av}^{\ast}$, defined above for the case $\Delta=2$ for
the two methods employed, i.e. the WL and BH methods. As discussed
above, the error bars quantify the large sample-to-sample
fluctuations of the specific heat peaks for the two methods
employed (note that the error bars with the larger cap-width
always refer to the WL method). From Fig.~\ref{fig1} it is
apparent that, while the sample mean averages
$[C_{m}^{\ast}]_{av}$ admits of finite-size scaling, the behavior
of $[C]_{av}^{\ast}$ appears as a random fluctuation around the
value $[C]_{av}^{\ast}\approx 0.815$, as shown by the dotted line
in this figure. In analogy with Fig.~\ref{fig1}, Fig.~\ref{fig2}
shows the corresponding susceptibility quantities, also for the
case $\Delta=2$. Again, the sample-to-sample fluctuations are very
large and the behavior of $[\chi]_{av}^{\ast}$ is clearly distinct
from that of $[\chi_{m}^{\ast}]_{av}$. Inspecting
Figs.~\ref{fig1},\ref{fig2} on a comparative basis we observe that
the deviations between the WL and BH methods are more pronounced
in the thermal case (Fig.~\ref{fig1}) and their difference is
noticeable for $L=20$. This difference represents the order of the
statistical errors of our scheme. Although these errors are still
small compared to the sample-to-sample fluctuations, they raise
doubts whether the total number of WL iterations $(j_{final}=20)$
is sufficient for the study of large lattice sizes.

From the discussion above, it is clear that when studying random
systems the only meaningful objects for investigating the
finite-size scaling behavior are the distributions of various
properties in ensembles of several realizations of the randomness.
Hence, it is important to be able to ascertain to what extent are
the results obtained from an ensemble of random realizations
representative of the general class to which the system belongs.
The answer hinges on the important issue of self-averaging. If a
quantity is not self-averaging, we talk about lack of
self-averaging and the process of increasing $L$ does not improve
the statistics. In other words, the sample-to-sample fluctuations
remain large. The problem of self-averaging in the 3D RFIM has
been a matter of investigation over the last
years~\cite{parisi02}. A common measure characterizing the
self-averaging property of a system based on the theory of
finite-size scaling has been discussed by Binder~\cite{binder88}
and has been used for the study of some random
systems~\cite{wiseman95,aharony96}. This measure inspects the
behavior of a normalized square width quantity, defined as:
\begin{equation}
\label{eq:11} R_{Q}=\frac{V_{Q}}{[Q]^{2}}
\end{equation}
where $V_{Q}=[Q^{2}]-[Q]^{2}$ is the sample-to-sample variance of
the average $[Q]$. Here, $Q$ is used in respect of the specific
heat $C$ and the susceptibility $\chi$. According to the
literature~\cite{binder88,wiseman95,aharony96} when the ratio
$R_{Q}$ tends to a constant, the system is said to be non
self-averaging and the corresponding distribution (say $P(Q)$)
does not become sharp in the thermodynamic limit.
\begin{figure}[htbp]
\includegraphics*[width=10 cm]{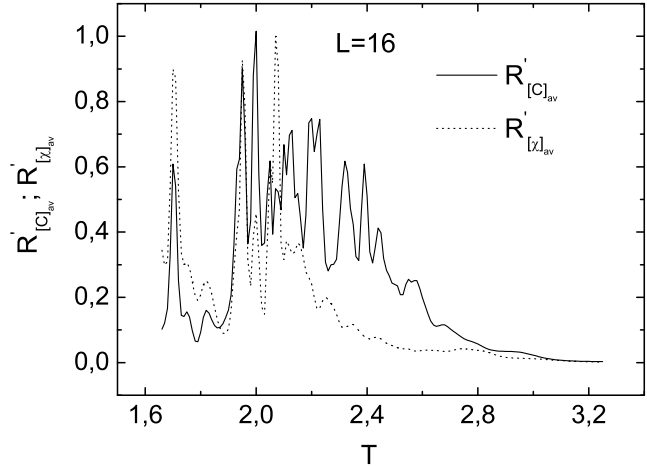}
\caption{\label{fig3}Temperature variation of the ratio
$R'_{[Q]_{av}}$ defined in the text, for both the specific heat
and susceptibility.}
\end{figure}

\begin{figure}[htbp]
\includegraphics*[width=10 cm]{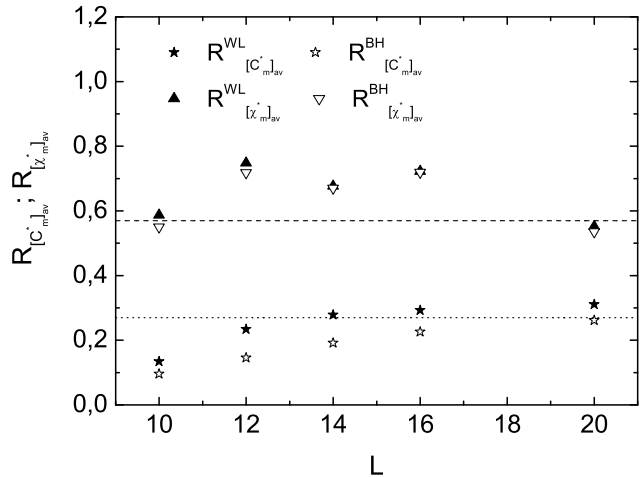}
\caption{\label{fig4}Illustration of the ratio
$R_{[Q_{m}^{\ast}]_{av}}$ of the specific heat and susceptibility,
for both the WL and BH methods, for randomness $\Delta=2$. Clear
saturation to a limiting non-zero constant value:
$R_{[C_{m}^{\ast}]_{av}}\rightarrow 0.28$ and
$R_{[\chi_{m}^{\ast}]_{av}}\rightarrow 0.57$.}
\end{figure}
Using our notation we may define the ratio $R_{Q}$ for the
specific heat and susceptibility in two explicit forms, one for
the case of the averaged curve $[Q]_{av}$:
\begin{equation}
\label{eq:12} R_{[C]_{av}}=\frac{V_{[C]_{av}}}{([C]_{av})^{2}},
\;R_{[\chi]_{av}}=\frac{V_{[\chi]_{av}}}{([\chi]_{av})^{2}}
\end{equation}
and one for the case of the average $[Q_{m}^{\ast}]_{av}$:
\begin{eqnarray}
\label{eq:13}
R_{[C_{m}^{\ast}]_{av}}=\frac{V_{[C_{m}^{\ast}]_{av}}}{([C_{m}^{\ast}]_{av})^{2}}=
\left(\frac{\sigma(C_{m}^{\ast})}{[C_{m}^{\ast}]_{av}}\right)^{2},\;
R_{[\chi_{m}^{\ast}]_{av}}=
\frac{V_{[\chi_{m}^{\ast}]_{av}}}{([\chi_{m}^{\ast}]_{av})^{2}}=
\left(\frac{\sigma(\chi_{m}^{\ast})}{[\chi_{m}^{\ast}]_{av}}\right)^{2}
\end{eqnarray}
In Fig.~\ref{fig3} we present the behavior of the ratio
$R'_{[Q]_{av}}=R_{[Q]_{av}}/R^{\ast}_{[Q]_{av}}$, where
$R^{\ast}_{[Q]_{av}}=\max\{R_{[Q]_{av}}\}$ as a function of the
temperature $T$, for $L=16$ and $\Delta=2$. The solid line
corresponds to the specific heat $(R'_{[C]_{av}})$ while the
dotted line to the susceptibility $(R'_{[\chi]_{av}})$. In this
figure only the results of the WL method are presented, since our
intention was to identify the temperature variation of the non
self-averaging property of the averaged specific heat and
susceptibility defined in Eq.~(\ref{eq:9}). Indeed, from
Fig.~\ref{fig3} we observe that for temperatures close to the
critical, the ratio $R_{[Q]_{av}}$ is maximized indicating
strongly non self-averaging behavior for both quantities. In
Fig.~\ref{fig4} we consider the behavior of the ratio
$R_{[Q_{m}^{\ast}]_{av}}$ of the specific heat
$(R_{[C_{m}^{\ast}]_{av}})$ and susceptibility
$(R_{[\chi_{m}^{\ast}]_{av}})$ as a function of the linear size
$L$, for both the WL and BH methods. Both ratios seem to tend to a
constant non-zero value, namely
$R_{[C_{m}^{\ast}]_{av}}\rightarrow 0.28$ and
$R_{[\chi_{m}^{\ast}]_{av}}\rightarrow 0.57$, confirming the above
mentioned lack of self-averaging of the specific heat and
susceptibility of the model.

\section{Summary and outlook}
\label{section4}

The numerical strategy applied in this paper enabled us to perform
extensive finite-temperature simulations and extract valuable
information for the generic behavior of the RFIM. It was shown
that the various definitions of the apparent finite-size anomalies
may not be equivalent. Our analysis revealed that the behavior of
the mean $[Q_{m}^{\ast}]_{av}$ is clearly distinct from that of
$[Q]_{av}^{\ast}$ and that this is directly connected to the lack
of self-averaging of the model. More work needs to be done towards
this direction, so that the subtle matter of self-averaging in the
RFIM is fully clarified and understood. One of our future plans,
is the verification of the above results by studying the model for
larger lattice sizes and various randomness values. In any case,
the present study puts forward some new ideas and an efficient
unified implementation of the DOS methods, suitable for the study
of random systems.

\begin{acknowledgments}
This research was supported by NKUA/SARG under Grant No.
$70/4/4071$.
\end{acknowledgments}

{}

\end{document}